\documentclass[journal]{IEEEtran}
\usepackage[dvips]{graphicx}
\usepackage{epsfig}
\usepackage{amsmath}
\usepackage{epsfig}
\usepackage{array}
\usepackage{amssymb}
\usepackage{color}

\newtheorem{Theorem}{\textbf{Theorem}}

\begin{document}

\title{A Remark on the AAA Method for Secret-Key Generation in Mobile Networks}

\author{Yingbo Hua, \emph{Fellow, IEEE}
\thanks{Department of Electrical and Computer Engineering,
University of California at Riverside, Riverside, CA 92521, USA. Email: yhua@ece.ucr.edu.
 This work was supported in part by the U.S. Department of Defense under W911NF-20-2-0267. The views and conclusions contained in this
document are those of the author and should not be interpreted as representing the official policies, either
expressed or implied, of  the U.S. Government. The U.S. Government is
authorized to reproduce and distribute reprints for Government purposes notwithstanding any copyright
notation herein.
}
}

\maketitle

\begin{abstract}
A broadly applicable method for secret-key generation is named for its accumulative, adaptable and additive (AAA) properties. This paper first shows a robustness of its performance. Namely, even if there is an inter correlation or a leakage caused intra correlation among the superimposed packets, provided
there is a nonzero probability for each packet to be missed in full or in part by Eve, then the equivocation of the key generated by the AAA method always becomes perfect as the number of superpositions becomes infinite. Also shown in this paper is a comparison between the AAA method and an ideal method based on reciprocal channel estimation, which reveals several advantages of the AAA method.
\end{abstract}

\begin{IEEEkeywords}
Secret-key generation, secret-key renewal, secret-key enhancement, mobile networks.
\end{IEEEkeywords}

\section{Introduction}
A recently studied method \cite{HUA2025109744} for secret-key generation between two or more mobile nodes across networks is based on packet superposition, but is named for its accumulative, adaptable and additive (AAA) properties.
Specifically, let $X_{l,i}$ be the $l$th bit randomly chosen according to a public protocol from the random payload (or its hashed form) of the $i$th packet shared between users.
So, $X_{l,i}$ and $X_{l',i}$ for $l\neq l'$ are independent to anyone who does not know the $i$th packet. Given $n$ packets at time $n$, an $L$-bits key can be generated by $K_{l,n}=X_{l,1}\oplus \cdots\oplus X_{l,n}$ (XOR) with $l=1,\cdots,L$, which is a most aggressive form of hash functions. If $X_{l,i}$ and $X_{l,i'}$ for all $i\neq i'$ are also independent, then the equivocation of the $L$-bits key $\mathcal{K}_n\doteq\{K_{1,n},\cdots,K_{L,n}\}$ at time $n$ against Eve can be shown \cite{HUA2025109744} to be $L$ (i.e., perfect) as long as one of the $n$ packets is missed by Eve.
This method is extremely simple and easy to implement. There is little latency caused by such a simple computation. There is no issue of key agreement since all bits used are already in agreement and authenticated by the parties. The key generation (or refreshing) rate can be as fast as the packet transmission rate between users, and there is no need for reciprocal channel or to wait for channel response to change. The generated or renewed secret key from the AAA method can be simply a by-product (for future use) of a normal process of packet transmissions between legitimate nodes. This is in great contrast against many reciprocal channel based methods developed in the past decades, e.g.,
see \cite{Hua2023Secure_Degree_of_Freedom}, \cite{Yang2024_10582429}, \cite{Cao2024_10603430}, \cite{Kojima2024_10495199}, \cite{Adil2024_10552259}, \cite{Tang2022_9697095} and many references therein.

In this paper, the performance of the AAA method is first revisited in section \ref{sec:performance} subject to possible dependencies among the superimposed packets and/or partial  leakage of each packet to Eve. It is shown that the secrecy achieved by the AAA method is still asymptotically perfect as the number of packets used becomes infinity.  Then a comparison between the AAA method and an ideal method based on a perfectly reciprocal channel gain is discussed in section \ref{sec:comparison}, which establishes several advantages of the AAA method over the other.

\section{Performance of the AAA Method}\label{sec:performance}
Let $E_{l,i}$ be the observation by Eve about $X_{l,i}$. Assume that the $i$th packet is intercepted by Eve with probability $1-\mu_i$, and missed by Eve with probability $\mu_i$. Then $E_{l,i}=X_{l,i}$ for all $l$ with probability $1-\mu_i$, and $E_{l,i}=\dag$ (erasure) for all $l$ with probability $\mu_i$.

The performance of the AAA method can be measured by the conditional entropy of $\mathcal{K}_n$:
\begin{equation}\label{}
  \epsilon_n^{(L)}\doteq \mathbb{H}(\mathcal{K}_n|\mathcal{E}^n)
\end{equation}
where $\mathcal{E}^n\doteq \{E_l^n;l=1,\cdots,L\}$ with $E_l^n=\{E_{l,i};i=1,\cdots,n\}$, and $\epsilon_n^{(L)}$ is also called the equivocation of $\mathcal{K}_n$ against Eve.

\subsection{Independent Packets}
If all $n$ packets are independent of each other, it can be shown \cite{HUA2025109744}  that $\epsilon_n^{(L)} = L\left (1-\prod_{i=1}^n(1-\mu_i)\right )$. In this case, $\epsilon_{n+1}^{(L)}>\epsilon_n^{(L)}$ and $\lim_{n\to\infty}\epsilon_n^{(L)}=L$ for $\mu_i>0$. Also note that $\epsilon_n^{(L)}=L$ if there is $i\in\{1,\cdots,n\}$ such that $\mu_i=0$. This means that the key generated by the AAA method at time $n$ has a perfect secrecy as long as Eve has missed at least one of the $n$ packets. In practice, all of the packets need to be encrypted by a secret key for authentication purpose. Even if all packets are intercepted by Eve and only one of them is encrypted by an initial secret key, the resulting key from the AAA method is still as secure as the initial secret key. For most mobile applications, it is highly unlikely for Eve to intercept all packets transmitted between two nodes over a time window where Eve's receive channel suffers a deep fade at least once.

\subsection{Dependent Packets}
Now we assume that there is a dependency among the packets used by the AAA method. Specifically, we assume that the $n$ packets follow a Markov model where $\texttt{Prob}(X_{l,i+1}=x_{l,i}|X_{l,i}=x_{l,i})=\alpha_{l,i}$, $\texttt{Prob}(X_{l,i+1}\neq x_{l,i}|X_{l,i}=x_{l,i})=1-\alpha_{l,i}$, and $X_{l,1}$ for all $l$ are independent and uniform.

\subsubsection{Asymptotical Analysis}
\begin{Theorem}
If $0<\alpha_{l,i}<1$ and $0<\mu_i\leq 1$ with $l=1,\cdots,L$ and $i=1,\cdots,n$, then $\lim_{n\to\infty}\epsilon_n^{(L)}=L$.
\end{Theorem}
\begin{IEEEproof}
Using the chain rule of entropy, we have $\epsilon_n^{(L)}=\mathbb{H}(K_{1,n}|\mathcal{E}^n)
  +\mathbb{H}(K_{2,n}|K_{1,n},\mathcal{E}^n)+\cdots$.
Recall $\mathcal{E}^n=\{E_1^n,\cdots,E_L^n\}$ where the $i$th entry of $E_l^n=\{E_{l,1},\cdots,E_{l,n}\}$ for each $l$ corresponds to either missed packet $i$ (i.e., having no information about $K_{1,n}$) or intercepted packet $i$ (i.e., having the same information about $K_{1,n}$ as the $i$ entry of $E_1^n$). Hence, $\mathbb{H}(K_{1,n}|\mathcal{E}^n)=\mathbb{H}(K_{1,n}|E_1^n)$.
For each realization of $\mathcal{E}^n$, $K_{1,n}$ is a sum of two groups of bits: $\mathcal{M}_1$ associated with missed packets, and $\mathcal{T}_1$ associated with intercepted packets; and similarly $K_{2,n}$ is a sum of $\mathcal{M}_2$ and $\mathcal{T}_2$. While $\mathcal{T}_1$ and $\mathcal{T}_2$ are one-to-one given this $\mathcal{E}^n$, $\mathcal{M}_1$ and $\mathcal{M}_2$ are independent. Hence, as $n\to\infty$, $\mathbb{H}(K_{2,n}|K_{1,n},\mathcal{E}^n)=\mathbb{H}(K_{2,n}|\mathcal{E}^n)
=\mathbb{H}(K_{2,n}|E_2^n)$.
The above analysis implies that
\begin{align}\label{}
  &\lim_{n\to\infty}\epsilon_n^{(L)}=\sum_{l=1}^L\lim_{n\to\infty}\mathbb{H}(K_{1,n}|E_l^n).
\end{align}

To prove $\lim_{n\to\infty}\mathbb{H}(K_{l,n}|E_l^n)=1$, let us write
\begin{equation}\label{eq:eln}
  \epsilon_{l,n} \doteq \mathbb{H}(K_{l,n}|E_l^n)
  =\sum_{e_l^n}p(e_l^n)h(\eta(e_l^n))
\end{equation}
where $e_l^n$ is any realization of $E_l^n$, $p(e_l^n)$ is the probability of $E_l^n=e_l^n$, $h(\eta)\doteq -\eta\log_2 \eta-(1-\eta)\log_2(1-\eta)$, and
\begin{equation}\label{}
  \eta(e_l^n)\doteq \texttt{Prob}(K_{l,n}\doteq X_{l,1}\oplus  \cdots \oplus X_{l,n}=0|E_l^n=e_l^n).
\end{equation}
Also define $\beta_{l,i}(e_l^n)\doteq \texttt{Prob}(X_{l,i}=1|E_l^n=e_l^n)$.

For simpler notations, we can now drop all the subscripts $l$ for the rest of the proof for $\epsilon_{l,n}$.
 It follows that
\begin{align}\label{}
  &\eta(e^n)=\sum_{m\texttt{ even }}\sum_{1\leq i_1<i_2<\cdots <i_m\leq n} \beta_{i_1}(e^n)\cdots\beta_{i_m}(e^n) \notag\\
  &\,\,\cdot \prod_{i\neq i_1,\cdots,i_m}(1-\beta_i(e^n)),
\end{align}
\begin{align}\label{}
  &1-\eta(e^n)=\sum_{m\texttt{ odd }}\sum_{1\leq i_1<i_2<\cdots <i_m\leq n} \beta_{i_1}(e^n)\cdots\beta_{i_m}(e^n) \notag\\
  &\,\,\cdot\prod_{i\neq i_1,\cdots,i_m}(1-\beta_i(e^n)).
\end{align}
Here $\eta(e^n)$ is  the probability that an even number of entries in $\{X_1,\cdots,X_n\}$ conditioned on $E^n=e^n$ are equal to ones. Similarly, $1-\eta(e^n)$ is the probability that an odd number of entries in $\{X_1,\cdots,X_n\}$ conditioned on $E^n=e^n$ are equal to ones.
It then follows that
\begin{align}\label{eq:1}
  &1=\eta(e^n)+(1-\eta(e^n))\notag\\
  &=\sum_{m=0}^n\sum_{1\leq i_1<i_2<\cdots <i_m\leq n} \beta_{i_1}(e^n)\cdots\beta_{i_m}(e^n)
  \notag\\
  &\,\,\cdot \prod_{i\neq i_1,\cdots,i_m}(1-\beta_i(e^n))\notag\\
  &=\prod_{i=0}^n(\beta_i(e^n)+(1-\beta_i(e^n))).
\end{align}
Furthermore, by reversing the sign of the first $\beta_i(e^n)$ for each $i$ in the above expression, we have
\begin{align}\label{eq:2}
&\delta(e^n)\doteq\prod_{i=0}^n(-\beta_i(e^n)+(1-\beta_i(e^n)))\notag\\
 & =\sum_{m=0}^n\sum_{1\leq i_1<i_2<\cdots <i_m\leq n} (-\beta_{i_1}(e^n))\cdots(-\beta_{i_m}(e^n))\notag\\
  &\,\,\cdot \prod_{i\neq i_1,\cdots,i_m}(1-\beta_i)\notag\\
  &=\eta(e^n)-(1-\eta(e^n)).
\end{align}
Combining \eqref{eq:1} and \eqref{eq:2}, we have $\eta(e^n)=\frac{1}{2}(1+\delta(e^n))$. Here
\begin{equation}\label{}
  \delta(e^n)=\prod_{i=0}^n(1-2\beta_i(e^n)).
\end{equation}

Since $h(\eta(e^n))$ equals $h(1-\eta(e^n))$, then $h(\eta(e^n))$ is invariant to any notational flipping on $K_n$ or equivalently on $X_i$. Hence without loss of generality, we can assume $0\leq \beta_i\leq \frac{1}{2}$ for all $i$.

Furthermore, we can partition the entire space $\{e^n\forall n\}$ into two parts: $\mathcal{S}_1$ and $\mathcal{S}_2$. For $e^n\in \mathcal{S}_1$, the number of positive (or nonzero) entries in $\{\beta_1(e^n),\cdots,\beta_n(e^n)\}$ \emph{increases without bound} as $n$ increases, i.e., the number of misses by Eve increases without bound as $n$ increases. For $e^n\in \mathcal{S}_2$, the number of positive (or nonzero) entries in $\{\beta_1(e^n),\cdots,\beta_n(e^n)\}$ \emph{does not increase without bound} as $n$ increases, i.e., the number of misses by Eve does not increase without bound as $n$ increases.

It follows from the definition of $\mathcal{S}_1$ and $\mathcal{S}_2$ that subject to $0<\alpha_i<1$ and $0<\mu_i\leq 1$ for all $i$,
\begin{equation}\label{}
  \lim_{n\to\infty}\sum_{e^n\in \mathcal{S}_1}p(e^n)=1,
\end{equation}
\begin{equation}\label{}
  \lim_{n\to\infty}\sum_{e^n\in \mathcal{S}_2}p(e^n)=0.
\end{equation}
Here $\mathcal{S}_1$ can be viewed as a typical subset of $\{e^n\forall n\}$ while $\mathcal{S}_2$ as an atypical subset of $\{e^n\forall n\}$. The reasoning behind the above two equations is that as $n$ increases, more and more entries of $\{X_1,\cdots,X_n\}$ will be missed by Eve, and hence more and more entries of $\{\beta_1(e^n),\cdots,\beta_n(e^n)\}$ will become positive (nonzero).

Therefore, for $e^n\in \mathcal{S}_1$, we have $\lim_{n\to\infty}\delta(e^n)=0$, $\lim_{n\to\infty}\eta(e^n)=\frac{1}{2}$, and hence $\lim_{n\to\infty}h(\eta(e^n))=1$.
Finally,
\begin{align}\label{}
  &\lim_{n\to\infty}\epsilon_n= \lim_{n\to\infty}\sum_{e^n\in \mathcal{S}_1}p(e^n)h(\eta(e^n))\notag\\
  &=\lim_{n\to\infty}\sum_{e^n\in \mathcal{S}_1}p(e^n)=1,
\end{align}
which completes the proof.
\end{IEEEproof}

\subsubsection{Discussion}
If $\mu_i=\mu$, $\alpha_{l,i}=\alpha$, $\mu'\doteq 1-\mu$ and $\alpha'\doteq 1-\alpha$, it can be shown \cite{HUA2025109744} that $\epsilon_{l,n}$ in \eqref{eq:eln} for $n=1,2,3$ are
\begin{equation}\label{}
  \epsilon_{l,1}=\epsilon_1=\mu,
\end{equation}
\begin{equation}\label{}
  \epsilon_{l,2}=\epsilon_2 = \mu^2+2\mu\mu' h(\alpha),
\end{equation}
\begin{align}\label{}
  &\epsilon_{l,3}=\epsilon_3 = \mu^3+2\mu\mu' h(\alpha)+\mu^2\mu' h(\alpha^2+\alpha'^2)\notag\\
  &\,\,+\mu\mu'^2\left [2\alpha\alpha'+(\alpha^2+\alpha'^2)h\left (\frac{\alpha^2}{\alpha^2+\alpha'^2}\right )\right ].
\end{align}
(There is a typo for $\epsilon_2$ in (14) of \cite{HUA2025109744} where $\mu_1(1-\mu_2)+(1-\mu_2)\mu_1$ should be $\mu_1(1-\mu_2)+(1-\mu_1)\mu_2$.)

But the derivation of $\epsilon_{l,n}$ for even a moderate $n>3$  is extremely tedious.
This is because the number of terms in the sum of \eqref{eq:eln} is the number of possibilities of $e_l^n$, which grows exponentially as $n$ increases. Even for a given $e_l^n$ and a moderate $n>3$, the computation of $\eta(e_l^n)$ is also not trivial in general. There is so far no closed form expression of $\epsilon_{l,n}$ for an arbitrary $n$, and the computation of $\epsilon_{l,n}$ is a hard problem as $n$ increases.
Any simulation attempt to compute $\epsilon_{l,n}$ encounters the same challenges.

\begin{figure}[!t]
  \centering
  \begin{minipage}[b]{0.48\linewidth}
    \centering
    \includegraphics[width=\linewidth]{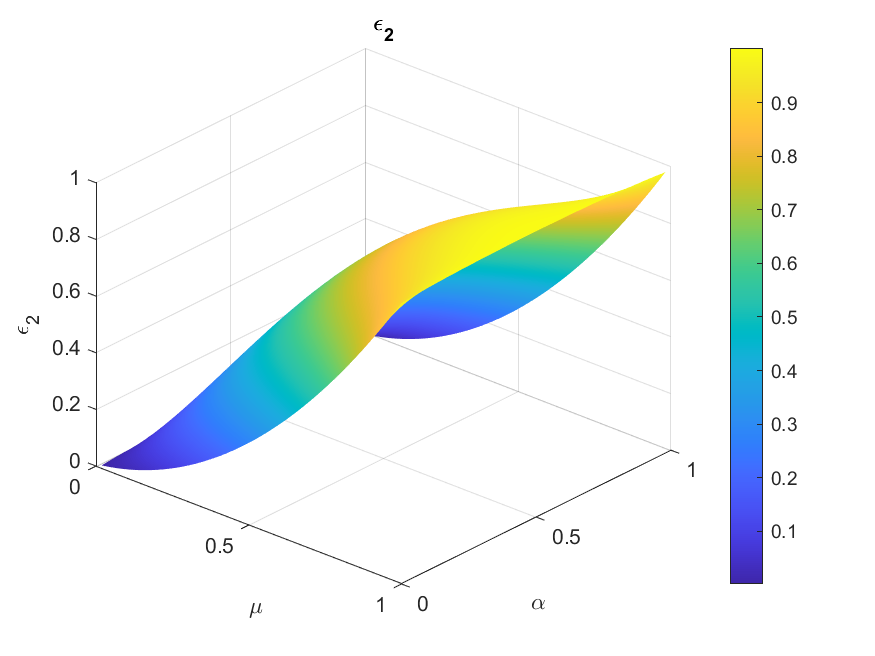}
   \caption{$\epsilon_2$ versus $\alpha$ and $\mu$.}\label{fig:epsilon2}
  \end{minipage}
  \hfill
  \begin{minipage}[b]{0.48\linewidth}
    \centering
    \includegraphics[width=\linewidth]{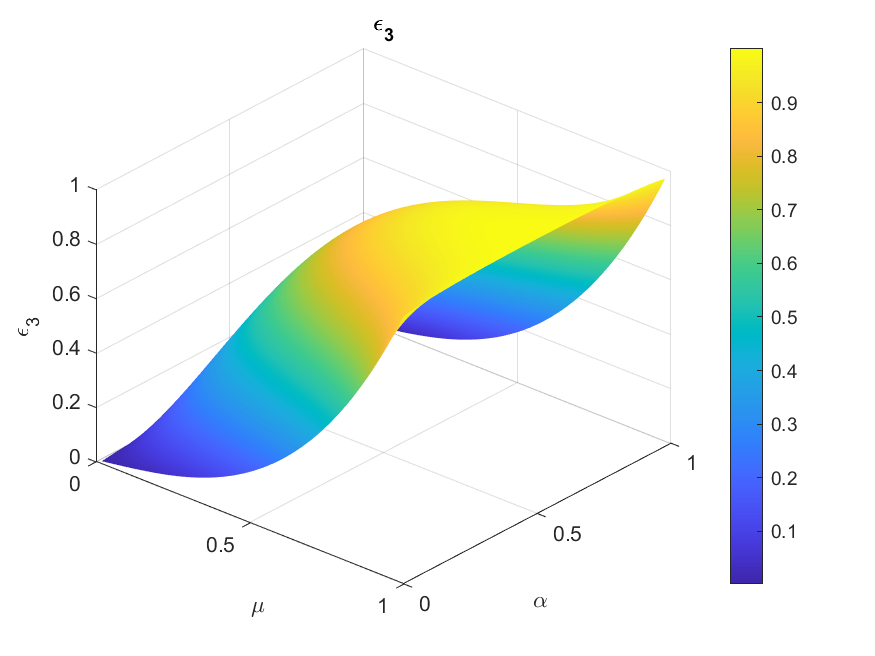}
    \caption{$\epsilon_3$ versus $\alpha$ and $\mu$.}\label{fig:epsilon3}
  \end{minipage}
\end{figure}
%
%

\begin{figure}[!t]
  \centering
  \begin{minipage}[b]{0.48\linewidth}
    \centering
    \includegraphics[width=\linewidth]{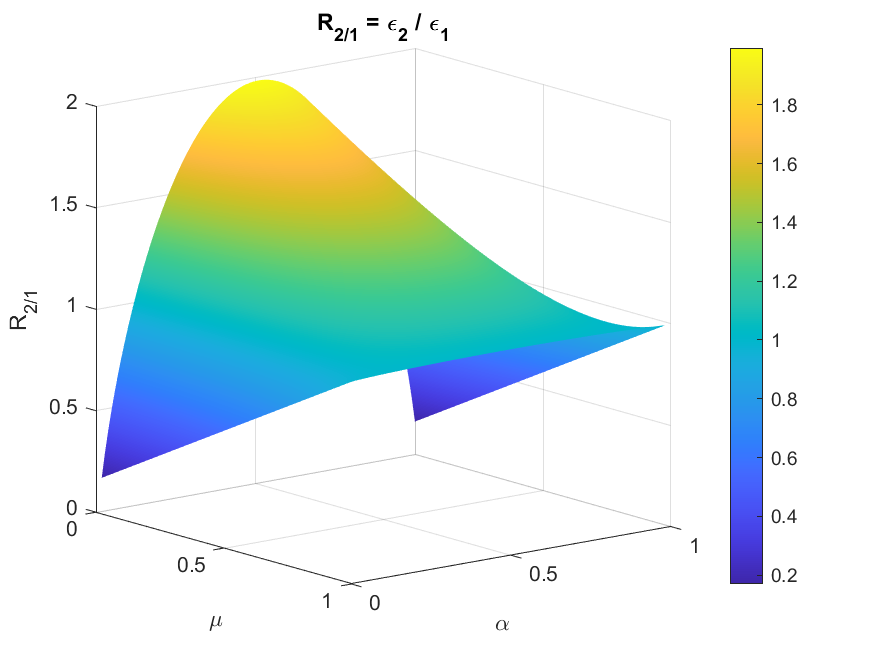}
    \caption{$\frac{\epsilon_2}{\epsilon_1}$ versus $\alpha$ and $\mu$.}\label{fig:R21}
  \end{minipage}
  \hfill
  \begin{minipage}[b]{0.48\linewidth}
    \centering
    \includegraphics[width=\linewidth]{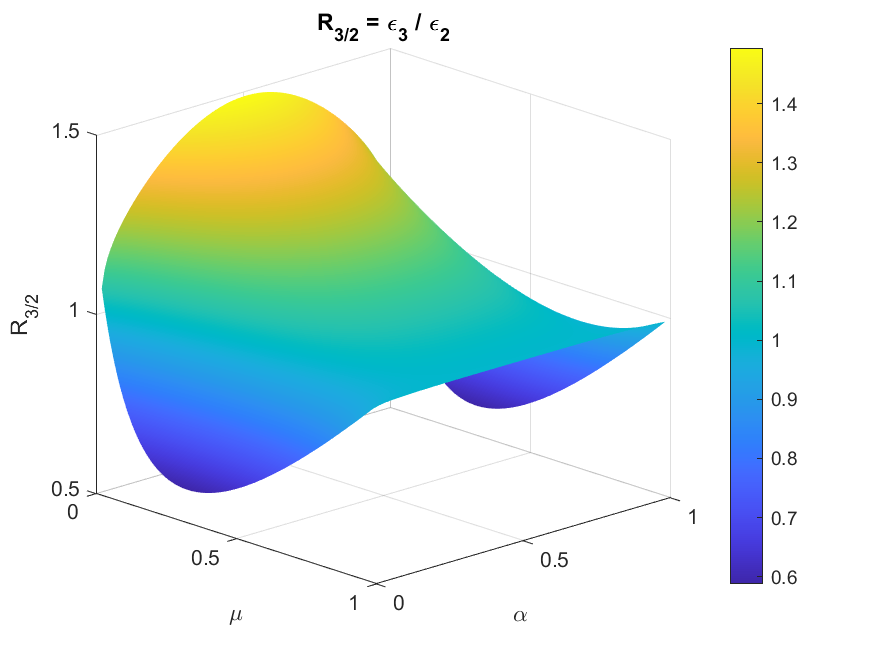}
   \caption{$\frac{\epsilon_3}{\epsilon_2}$ versus $\alpha$ and $\mu$.}\label{fig:R32}
  \end{minipage}
\end{figure}

The above $\epsilon_2$ and $\epsilon_3$ versus $\alpha$ and $\mu$ are illustrated in Figs. \ref{fig:epsilon2} and \ref{fig:epsilon3}, which are nonlinear functions of $\alpha$ and $\mu$. The ratios $\frac{\epsilon_2}{\epsilon_1}$ and $\frac{\epsilon_3}{\epsilon_2}$
 versus $\alpha$ and $\mu$ are illustrated in Figs. \ref{fig:R21} and \ref{fig:R32}.
Here we see that $\epsilon_n$ is not a monotonic function of $n$ in general unless $\alpha$ is close enough to $\frac{1}{2}$. But we also see from Fig. \ref{fig:epsilon3} that $\epsilon_3$ tends to bulge toward one over a broader area of $\alpha$ and $\mu$ than $\epsilon_2$ does, which supports the proven theorem.

%
%


\subsection{Packets with Partial Leakage to Eve}
More generally, when a packet is transmitted wirelessly from user A to user B, part of its information could be intercepted by Eve. The secrecy capacity in bits/s/Hz for the $i$th packet is known to be $C_{i,s}=(C_{i,U}-C_{i,E})^+$, with $C_{i,U}$ being the capacity  between users, and $C_{i,E}$ being the capacity from user A to Eve. But here assume that users apply a linear block code such as LDPC code to reliably transmit a message of $L$ (uniformly random) information bits  in the $i$th packet, denoted by the $L\times 1$ vector $\mathbf{m}_i$ in the binary field $\mathbb{F}_2^{L\times 1}$. Then in theory, Eve may use the same code to determine $\mathbf{m}_i$ up to $0\leq l_i\leq L$ unknown bits. More specifically, Eve knows that after her optimal error correction, there is a matrix $\mathbf{A}_i\in \mathbb{F}_2^{L\times l_i}$ such that the error vector in the message
 received by Eve can be expressed as
\begin{equation}\label{eq:mAa}
  \Delta \mathbf{m}_i=\mathbf{A}_i\mathbf{a}_i
\end{equation}
where $\mathbf{a}_i\in\mathbb{F}_2^{l_i\times 1}$ is in complete secrecy against Eve. Furthermore, $\mathbf{A}_i$ is dependent on the channel noise at Eve associated with the $i$th packet, and
$l_i\geq L- n_sC_{i,E}$ with $n_s$ being the number of symbols (channel uses) exploiting the code.

In this case, the key generated by the AAA method at users after $n$ successfully exchanged packets  can be expressed by $\mathbf{k}^{(n)}=\oplus_{i=1}^n\mathbf{m}_i\in\mathbb{F}_2^{L\times 1}$. But the error vector in the key generated by Eve using the same method is
\begin{equation}\label{}
  \Delta \mathbf{k}^{(n)}=\mathbf{A}^{(n)}\mathbf{a}^{(n)}
\end{equation}
where $\mathbf{A}^{(n)}=[\mathbf{A}_1,\cdots,\mathbf{A}_n]$ and $\mathbf{a}^{(n)}=[\mathbf{a}_1^T,\cdots,\mathbf{a}_n^T]^T$. Let $r_n$ be the rank of the discrete matrix $\mathbf{A}^{(n)}\in\mathbb{F}_2^{L\times \sum_{i=1}^nl_i}$. Then the secrecy or equivocation of $\mathbf{k}^{(n)}$ against Eve is (at least) $r_n$.
 Due to random noises at Eve, there is a large probability  that $r_n=\min(L,\sum_{i=1}^nl_i)$. So $r_n$ tends to increase with $n$ until $r_n=L$. Let $p_n$ be the probability of $r_n=L$, and assuming a positive probability for $l_i\geq 1$ due to Eve's channel fading, one can expect $\lim_{n\to\infty}  p_n=1$.

Also note that in practice it is generally still hard for Eve to find $\mathbf{A}_i$ which is noise dependent. So, in the next section, we assume that Eve either knows the complete $\mathbf{m}_i$ (corresponding to an intercepted packet) or nothing about $\mathbf{m}_i$ (corresponding to a missed packet).

\section{Comparison with Reciprocal Channel Estimation}\label{sec:comparison}

Next we show a comparison between the AAA method and a method based on a perfectly reciprocal channel gain (RCG). The latter highlights a fundamental essence of physical-layer key generation in such works as in \cite{Yang2024_10582429}, \cite{Cao2024_10603430}, \cite{Kojima2024_10495199}, \cite{Adil2024_10552259}, \cite{Tang2022_9697095} and many others. Note that the following comparison focuses on primary physical-layer costs of both methods but ignores the communication overhead and latency associated with information reconciliation required by the RCG method.

\subsection{Using reciprocal channel estimation}
Assume a pair of reciprocal fading channels between two users as follows:
\begin{equation}\label{eq:yml}
  y_{m,s}=\sqrt{p} h_m x_{m,s}+ w_{m,s},
\end{equation}
\begin{equation}\label{eq:yml_p}
  y_{m,s}'=\sqrt{p} h_m x_{m,s}'+ w_{m,s}',
\end{equation}
where $\sqrt{p}x_{m,s}$ is the signal (of power $p$) transmitted by user A in the first half of the $m$th channel coherence period, $\sqrt{p}x_{m,s}'$ is the signal (of power $p$) transmitted by user B in the second half of the $m$th channel coherence period, and $s=1,\cdots,S$ denotes the sampling time (or symbol index) in a given time window. So, each channel coherence period is no less than $2S$ symbol intervals. For reciprocal-channel based key generation, let us assume $x_{m,s}$ and $x_{m,s}'$  to be public pilots, which could be identical to each other. But the noises $w_{m,s}$ and $w_{m,s}'$ for all $m$ and $s$ are modelled to be i.i.d. circular complex Gaussian, i.e., $\mathcal{CN}(0,1)$.

Here $h_m$ is the same channel response parameter in both directions (i.e., perfectly reciprocal) in the $m$th coherence period. Also assume that $h_m$ for $m=1,\cdots,M$ are i.i.d. $\mathcal{CN}(0,1)$.

Corresponding to $x_{m,s}$ and $x_{m,s}'$ transmitted by users A and B respectively, the signals received by Eve are
\begin{equation}\label{eq:zml}
  z_{m,s}=\sqrt{p}\gamma_m g_m x_{m,s}+v_{m,s},
\end{equation}
\begin{equation}\label{eq:zml_p}
  z_{m,s}'=\sqrt{p}\gamma_m' g_m x_{m,s}'+v_{m,s}',
\end{equation}
where $\gamma_m$ and $\gamma_m'$ account for the large scale factors between user A and Eve and between user B and Eve in the $m$th coherence period. Here $v_{m,s}$ and $v_{m,s}'$ for all $m$ and $s$ are i.i.d. $\mathcal{CN}(0,1)$, and so are $g_m$ for all $m$.

  As an ideal condition for key generation using channel estimation, we assume  that $g_m$ and $h_m$ are independent of each other, which is however not always the case in practice.

For a compact expression, let us rewrite \eqref{eq:yml} as
\begin{equation}\label{}
  \mathbf{y}_m=\sqrt{p}h_m\mathbf{x}_m+\mathbf{w}_m,
\end{equation}
where $\mathbf{y}_m = [y_{m,1},\cdots,y_{m,S}]^T$, and $\mathbf{x}_m$ and $\mathbf{w}_m$ are similarly defined.
The MMSE estimate of $h_m$ by user B from $\mathbf{y}_m$ can be written as
\begin{equation}\label{eq:hathk}
  \hat h_m = \sqrt{p}\mathbf{x}_m^H(p\mathbf{x}_m\mathbf{x}_m^H+\mathbf{I})^{-1}\mathbf{y}_m
  =\frac{\sqrt{p}}{Sp+1}\mathbf{x}_m^H\mathbf{y}_m,
\end{equation}
and the MSE of $\hat h_m$
is
\begin{align}\label{}
  &\sigma_{\Delta h}^2 \doteq\mathbb{E}\{|\hat h_m-h_m|^2\}= 1-\sqrt{p}\mathbf{x}_m^H(p\mathbf{x}_m\mathbf{x}_m^H+\mathbf{I})^{-1}
  \sqrt{p}\mathbf{x}_m\notag\\
  &
  =1-Sp/(Sp+1)
  =1/(Sp+1).
\end{align}
Note that $\sigma_{\hat h}^2 \doteq\mathbb{E}\{|\hat h_m|^2\}=\frac{p}{(Sp+1)^2}\mathbf{x}_m^H(p
\mathbf{x}_m\mathbf{x}_m^H+\mathbf{I})\mathbf{x}_m
=\frac{Sp}{Sp+1}$.

Similarly, \eqref{eq:yml_p} can rewritten as
\begin{equation}\label{}
  \mathbf{y}_m'=\sqrt{p}h_m\mathbf{x}_m+\mathbf{w}_m',
\end{equation}
and the MMSE estimate of $h_m$ by user A from $\mathbf{y}_m'$ is
\begin{equation}\label{}
  \hat h_m' = \sqrt{p}\mathbf{x}_m^H\mathbf{y}_m'/(Sp+1),
\end{equation}
where we have used $x_{m,s}=x_{m,s}'$.

It follows that the secret-key capacity (in bits per coherence period) achievable by users A and B is
\begin{align}\label{}
  &C_1\doteq \mathbb{I}(\mathbf{y}_m;\mathbf{y}_m')=\mathbb{I}(\hat h_m;\hat h_m')
  =\texttt{h}(\hat h_m)-\texttt{h}(\hat h_m|\hat h_m')\notag\\
  &
  =\log_2\sigma_{\hat h}^2-\log_2\bar \sigma_{\Delta h}^2
\end{align}
where $\mathbb{I}(a;b)$ denotes mutual information between $a$ and $b$,  $\texttt{h}(\cdot)$ denotes differential entropy, $\bar \sigma_{\Delta h}^2$ is the MSE of the MMSE estimate of $\hat h_m$ from $\hat h_m'$. It follows that
\begin{equation}\label{}
  \bar \sigma_{\Delta h}^2=\sigma_{\hat h}^2-(\mathbb{E}\{\hat h_m\hat h_m'^H\})^2(\mathbb{E}\{\hat h_m'\hat h_m'^H\})^{-1}.
\end{equation}
Here
\begin{align}\label{}
  &\mathbb{E}\{\hat h_m\hat h_m'^H\}
  =\mathbb{E}\left \{\frac{p}{(Sp+1)^2}\mathbf{x}_m^H\mathbf{y}_m\mathbf{y}_m'^H\mathbf{x}_m\right\}
  \notag\\
  &=\frac{p^2}{(Sp+1)^2}\mathbf{x}_m^H\mathbf{x}_m\mathbf{x}_m^H\mathbf{x}_m
  =\frac{S^2p^2}{(Sp+1)^2},
\end{align}
\begin{align}\label{}
  &\mathbb{E}\{\hat h_m'\hat h_m'^H\}
  =\mathbb{E}\left \{\frac{p}{(Sp+1)^2}\mathbf{x}_m^H\mathbf{y}_m'\mathbf{y}_m'^H\mathbf{x}_m\right\}
  \notag\\
  &
  =\frac{p}{(Sp+1)^2}\mathbf{x}_m^H(p\mathbf{x}_m\mathbf{x}_m^H+\mathbf{I})\mathbf{x}_m
  =\frac{Sp}{Sp+1}.
\end{align}
Hence
$
  \bar \sigma_{\Delta h}^2=\frac{Sp}{Sp+1}-\frac{S^2p^2}{(Sp+1)^2}\frac{Sp}{Sp+1}
  =\frac{Sp(2Sp+1)}{(Sp+1)^3}
$.
Then
\begin{align}\label{}
  &C_1=\log_2\frac{(Sp+1)^2}{(2Sp+1)}
  =\log_2\left (1+\frac{S^2p^2}{2Sp+1}\right )\notag\\
  &
  <\log_2\left (1+\frac{1}{2}Sp\right )
\end{align}
where the inequality is also an approximation if $2Sp\gg 1$. Recall that $C_1$ is in bits per coherence period. With $M$ coherence periods, the maximum achievable number $L_1$ of (independent) bits in the secret key is
\begin{equation}\label{eq:L1}
  L_1=\gamma MC_1< \gamma M\log_2\left (1+\frac{1}{2}Sp\right ),
\end{equation}
which is subject to a sufficient amount of computations and communications needed for quantization, information reconciliation and privacy amplification.
Here $\gamma<1$ measures the efficiency of the key generation. In theory, as $M\to\infty$,  then $\gamma\to 1$, and the equivocation of this key of size $L_1$ becomes perfect.

\subsection{Using the AAA method}

In this case, we assume that user A transmits to user B a packet consisting of $S$ pilot symbols and $S$ information symbols, where every symbol has the power $p$. Assume a large $S$ so that $Sp\gg 1$. Then the channel estimation error (of MSE $\frac{1}{Sp+1}$) at user B using the pilot symbols causes negligible effect on the detection of the information symbols since the channel noise has the unit variance. Then the capacity of the information (in bits per information symbol) transmitted from user A to user B could be approximately up to
\begin{equation}\label{}
  C_{m,U}=\log_2(1+p|h_m|^2)
\end{equation}
where $h_m$ is the receive channel response at user B.

But $h_m$ is unknown to user A (the transmitter) for each coherence period. So, user A can only choose a predetermined rate $R$ (in bits per information sample) for the transmitted packet. This packet can be successfully detected by user B if $R<C_{m,U}$. On the other hand, if $R>C_{m,U}$, this packet is treated as lost. The packet loss rate at user B is
\begin{align}\label{}
  &\mu_U\doteq \texttt{Prob}(R>C_{m,U})=\int_0^{\frac{2^R-1}{p}}e^{-x}dx\notag\\
  &=1-\exp\left (-\frac{2^R-1}{p}\right )
\end{align}
where the PDF of $|h_m|^2$, i.e.,  $e^{-x}$ for $x>0$, has been used. If  $\frac{2^R-1}{p}\ll 1$ (high power), then
$\mu_U\approx \frac{2^R-1}{p}$.

Similarly, the packet loss rate at Eve subject to the packet being received by user B can be in many applications modelled as
\begin{align}\label{}
  &\mu_{E,m}\doteq \texttt{Prob}(R>C_{m,E}|R<C_{m,U})=\texttt{Prob}(R>C_{m,E})\notag\\
  &
  =1-\exp\left (-\frac{2^R-1}{p\gamma_m^2}\right ),
\end{align}
where $C_{m,E}=\log_2(1+p\gamma_m^2|g_m|^2)$ based on \eqref{eq:zml} has been applied.

So, a predetermined (or raw) key size chosen by the AAA method is $SR$ (in bits). After $M$ transmissions over $M$ coherence periods, the effective raw key size becomes $SR(1-\mu_U^M)$, which accounts for the probability that all packets from user A to user B are lost.
The equivocation of each bit in the key  is
\begin{equation}\label{}
  \epsilon_M= 1-\prod_{m=1}^M(1-\mu_{E,m}),
\end{equation}
which is similar to a discussion shown earlier based on independent packets.
Hence the final effective key size is
\begin{equation}\label{eq:L2}
  L_2 = SR(1-\mu_U^M)\left (1-\prod_{m=1}^M(1-\mu_{E,m})\right ).
\end{equation}
Typically, $\mu_U\ll 1$ and hence $L_2 \approx SR\left (1-\prod_{m=1}^M(1-\mu_{E,m})\right )$. For example, if $\mu_{E,m}=0.1$ and $M=22$, then $L_2\approx 0.9SR$.

\subsection{Comparison between $L_1$ and $L_2$}

It follows from \eqref{eq:L1} and \eqref{eq:L2} that
\begin{itemize}
  \item $L_1$ increases linearly with $M$. But a large $M$ means a large latency, which in practice imposes an upper limit on $M$. At the same time, $L_1$ is only proportional to $\log_2 S$ for a large $S$.
  \item $L_2$ increases linearly with $S$, and converges to $SR$ very quickly (if not all of $\mu_{E,m}$ are very small) as $M$ increases. In fact, $L_2$ becomes $SR$ if there is $m\in\{1,\cdots,M\}$ such that $\mu_{E,m}=1$.
\end{itemize}

So, in practice, if $SR$ is sufficiently long for a secret key, and/or there is such a latency constraint on $M$ that $M<SR$ but $\prod_{m=1}^M(1-\mu_{E,m})\ll 1$, then the AAA method is preferred.

On the other hand, if $SR$ is a too short length for a secret key, and $M$ can be so large that $M>SR$, then the use of reciprocal channel for key generation could be justified. However, in this case, the AAA method can choose to cascade multiple packets transmitted over multiple coherence periods to form a secret key of larger size. In this way, the AAA method can be still superior to the reciprocal-channel based methods in terms of key size (or security) and simplicity (or practicality).

More importantly, the AAA method does not require channel reciprocity, channel gain variations, or  any transmission solely for the purpose of secret-key generation. The AAA method only needs packets that are already transmitted and authenticated between users. This is a superior advantage of the AAA method over those in many prior works that require reciprocal and variable channels and additional transmissions for channel probing and/or information reconciliation. The only crucial requirement by the AAA method is that there is a non-negligible packet loss rate at Eve. For most terrestrial networks where no Eve is able to intercept all packets exchanged between legitimate users, the AAA method seems the best option to solve the problem of secret-key generation and/or renewal.

For WiFi, the payload in a packet can be 12,000 bits, and a typical channel coherence time is somewhere between 1-200ms (depending on the mobility of the environment). For LoRa, a typical payload in a packet is 88–1936 bits (due to low transmission power), and a typical channel coherence time is between 10-300ms. For ZigBee, they are respectively 640-800 bits, and 50–500ms. A secret key of virtually any size needed in such applications can be generated or refreshed in real time by the AAA method (even before using any concatenation).

\section{Conclusion}
This paper has revisited a recently studied method (called the AAA method) for secret-key generation. It is shown that the AAA method can asymptotically yield a perfect secrecy of its generated key of a predetermined size even if the data packets used for superposition are either inter or intra (caused by leakage) correlated. Also shown is a comparison between the AAA method and an ideal method for secret-key generation based on perfectly reciprocal channel gain. As long as the packet loss rate at Eve is not very small during a time window of interest, the AAA method shows a great advantage over the other method. This paper has provided additional fundamental insights into the AAA method. Experimental works on the AAA method are highly desirable in the future, which should include a careful study of realistic packet loss rate at Eve for WiFi, ZigBee, LoRa and other mobile applications. But for applications where Eve is able to intercept all packets of interest (such as in satellite communications where Eve is within the same transmission beamwidth for a legitimate receiver), secret-key generation via radio communications should focus on the exploitation of the \emph{finite} signal-to-noise ratio at Eve's receiver \cite{Hua_Maksud_2024} and \cite{STEEP}.

\bibliographystyle{IEEEtran}
\bibliography{references}

\end{document}